\documentclass{amsart}
\usepackage{amssymb,amsmath,amsfonts,latexsym,amsthm,mathrsfs}
\usepackage{amsaddr}
\usepackage{graphicx}
\numberwithin{equation}{section}

\begin{document} 
	\title{Bounds on the complex viscoelasticity \\ for
		surface waves on ice-covered seas}
	\author{C. Sampson$^1$, D. Hallman$^2$, N. B. Murphy$^2$, E. Cherkaev$^2$, and K. M. Golden$^2$}
	\address{$^1$Joint Center For Satellite Data Assimilation, UCAR \\ $^2$Department of Mathematics, University of Utah}
	\maketitle

\begin{abstract}
Oceanic wave propagation through Earth's sea ice covers is a critical component of accurate ice and climate modeling. Continuum models of the polar ocean surface layer are characterized rheologically by the effective complex viscoelasticity of the composite of ice floes and sea water. Here we present the first rigorous theory of this parameter, and distill its dependence on mixture geometry into the spectral properties of a self-adjoint operator analogous to the Hamiltonian in quantum physics. Bounds for the complex viscoelasticity are obtained from the sea ice concentration and the contrast between the elastic and viscous properties of the ice and water/slush constituents. We find that several published wave attenuation datasets in both laboratory and field settings fall well within the bounds for specific contrast values of the ice/ocean composite.  
\end{abstract}

\section{Introduction}
The interaction of ocean surface waves with Earth's sea ice covers
is a fundamental process impacting the dynamics and thermodynamics
of the polar marine environment. 
As wave energy propagates through the 
upper layer of the ocean, a composite of sea ice and sea water,
it can break ice floes 
and alter the floe size
distribution and sea ice concentration field.
Wave-ice interactions 
can influence atmosphere-ocean fluxes and 
melting and freezing processes, 
accelerating melting by breaking larger floes or 
promoting pancake ice formation, which impacts
crystalline structure
\cite{Squire2018,Roach_JAMES_2019,Golden_NAMS_2020,bennetts2022}.

Perhaps the most dynamic component of the Arctic sea ice cover is the 
marginal ice zone (MIZ), the transitional region between
dense pack ice to the north and open ocean to the south.
It is a biologically active region characterized by 
intense air-ice-ocean and wave-ice interactions.
Formation of the MIZ results from numerous processes 
including ice breakage by ocean waves and turbulent 
thermodynamic mixing. Ultimately, the MIZ serves as a 
physical buffer around pack ice and plays a fundamental 
role in climate and polar ecosystems, with wave penetration 
being a defining property of MIZ extent
\cite{massom_polar_2006,bennetts2022,Strong_SR_2024}.

In recent years, there has been an increasing realization of the importance of ocean waves in the growth and decay of the seasonal ice pack. In fact, a striking correlation between Antarctic sea ice extent and wave activity has been found 
\cite{Kohout2014} and the role of oceanic waves in catastrophic ice shelf collapse has been investigated \cite{Massom2018}. In any ice-covered region, the area fraction of ocean surface covered by ice is known as the sea ice concentration $\psi$, a standard satellite data product. The sea ice concentration $\psi$ and the ice floe size distribution
play a significant role in wave propagation characteristics.
Ocean waves break up and shape ice floe geometry, 
which in turn, controls which amplitudes and wavelengths propagate further into the pack. 
Ice-ocean interactions have become increasingly important in the Arctic with the precipitous declines of summer ice extent and increases in wave activity \cite{Waseda2018},
which occur in conjunction with a widening MIZ \cite{Strong2013,Strong_SR_2024}. These recent changes 
have complex, dynamic implications for both sea ice formation in winter and melting in summer \cite{jmse9040365}. 
A survey of the history and questions at the forefront of the field may be found in \cite{Squire2018,Roach_JAMES_2019}.

Continuum models have been developed which treat the interface between the atmosphere 
and ocean
as a relatively thin layer made of a two component composite of sea ice and slushy water. Several of the proposed models treat the ice and slushy water mix as a single material atop an inviscid ocean. These models are particularly appropriate for waves with long wavelength compared to the floe size. The top layer has been modeled to be purely elastic \cite{Bates1980}, purely viscous \cite{Keller1998}, and viscoelastic \cite{Mosig2015, Wang2010}. Models which incorporate the flexibility of an additional layer include the viscous models of \cite{DECAROLIS2002399, SUTHERLAND2019111} which use highly viscous and impermeable layers, and \cite{CHEN201988} which uses a poroelastic layer and a weakly compressible fluid layer. At the heart of these models are analytically derived dispersion relations which feature rheological parameters, namely, the effective elasticity, viscosity, and/or complex viscoelasticity. These parameters control an energy damping shown to be critical to accurate representation of wave propagation in the MIZ \cite{Meylan2021}. The effective parameters, which are influenced by ice-ocean composite geometry and the physical properties of the constituents, are difficult to determine analytically and are typically fitted to models through measured attenuation data.


In this study we introduce a novel approach to the study of wave attenuation in sea ice. We are motivated by the theory of the effective complex permittivity for electromagnetic wave propagation through two-phase composites, and employ the analytic continuation method (ACM) of homogenization
\cite{Bergman:PRL-1285,Milton:APL-300,Golden1983,milton2002,Cherkaev:JBiomech-345,Cherkaev:2019}.
We find a Stieltjes integral representation for the complex viscoelasticity (also called the complex shear modulus) of the two layer model presented in \cite{Wang2010} in a quasistatic setting. The model of Wang and Shen has experienced wide use and has been incorporated into Wave Watch III \cite{WWIII}. We choose this model as it is sufficiently general and can be reduced to the purely viscous or purely elastic cases. Our approach is based on earlier work \cite{Cherkaev:JBiomech-345,Cherkaev:2019} where a Stieltjes integral representation is derived for the effective complex viscoelastic modulus of two-phase compressible viscoelastic materials. In the past, the integral representation has been extended to effective elastic properties \cite{Bruno1993,dell1986,milton2002,Ou2006}. A Stieltjes 
representation for the effective viscoelastic shear modulus has also been obtained for the case of torsion of a viscoelastic cylinder whose microstructure is uniform in the axial direction \cite{Lista2007,Lista2008,tok2001}. 
We note that this Stieltjes approach has also been developed recently to obtain rigorous
bounds on the thermal conductivity of sea ice in the presence of 
convective fluid flow \cite{Kraitzman_PRSA_2024}.

The integral representation we find in this work involves the 
spectral measure of a self-adjoint operator which 
provides bounds on the effective viscoelastic parameter of the floating ice cover modeled as an incompressible ice-slush composite material. 
The bounds themselves depend on the moments of the
measure: the more moments are known, the tighter the bounds.
In this current work, we focus on bounds produced by the zeroth moment of the spectral measure which is precisely equal to the sea ice concentration $\psi$.



















\section{Methods}

\subsection{Dispersion relation}
In \cite{Wang2010}, the ice cover is modeled as a homogeneous, isotropic Kelvin-Voight material with finite thickness overlying an inviscid layer (the ocean). The homogeneous ice layer is assumed to be incompressible, and so its displacement $u$ satisfies $\nabla \cdot u = 0$. The deformation of the layer due to wave motion is described by a strain tensor $\boldsymbol{\epsilon}$, where $\boldsymbol{\epsilon}=(1/2) (\nabla u +\nabla^T  u)=\nabla^S u$ is the symmetric gradient of the displacement $u$. The stress and strain tensors can be decomposed into volumetric and deviatoric components
\begin{align}
	\boldsymbol{\sigma} = \boldsymbol{\sigma}_v + \boldsymbol{\sigma}_s, \quad \boldsymbol{\epsilon} = \boldsymbol{\epsilon}_v + \boldsymbol{\epsilon}_s.
\end{align}
The volumetric and deviatoric components correspond to shape-preserving and volume-preserving deformations, respectively. 

For a Kelvin-Voigt material, the relationship between the deviatoric stress and strain is given by $\boldsymbol{\sigma}_s=2G\boldsymbol{\epsilon}_s+2\upsilon \rho \boldsymbol{\dot \epsilon}_s$ where $G$ is the shear modulus,  $\upsilon$ is the kinematic viscosity, and $\rho $ is the density.  
The momentum equation for the ice layer is 
\begin{equation}
	\label{momentum}
	\rho \ddot{u}=\nabla \cdot (\boldsymbol{\sigma}_s +P\boldsymbol{I})-\rho  g e_z\,,
\end{equation}
where $P$ denotes pressure, $\boldsymbol{I}$ is the identity matrix, $g$ is gravitational acceleration, and $e_z$ is the $z$-direction unit vector. We remark here that the equations of motion \eqref{momentum} are given in terms of the displacement field $u$ as opposed to the velocity field $\dot u$ described in \cite{Wang2010}. 

The dispersion relation developed in \cite{Wang2010} is concisely written in \cite{Mosig2015} as
\begin{align} \label{dispersion}
		D_{\text{WS}}(Q_{\text{WS}}gk \tanh(H k) - \omega^2) = 0 \,,
\end{align}
where $H$ is the depth of the water layer, $\omega$ is the angular frequency of the oceanic wave, and $k=k_r + ik_i$ is the complex wavenumber containing both the spatial oscillation $k_r$ as well as the wave amplitude attenuation coefficient $k_i$. We refer the reader to \cite{Mosig2015} for the full expressions of the terms $D_{\text{WS}}(k,\omega)$ and $Q_{\text{WS}}(k,\omega)$, as they are lengthy and hence omitted here. We consider the \emph{hydrostatic/quasistatic limit} where $\omega<<1$ and $|k|<<1$ in which the dispersion relation (\ref{dispersion}) simplifies considerably to
\begin{align} \label{SDP}
		\nu^* = \frac{\rho}{4}\bigg(\frac{\omega}{k}\bigg)^2,
\end{align}
where $\nu^*$ is the effective complex viscoelasticity of the homogeneous ice layer. Using a harmonic wave profile $u=A \exp(kx-i \omega t)$ with amplitude $A$ leads to the relations $\boldsymbol{\dot \epsilon}_s=-i\omega \boldsymbol{\epsilon}_s$ and $\boldsymbol{\sigma}_s=2\nu\boldsymbol{\epsilon}_s$. Here, $\nu=G-i\omega \rho \upsilon$ represents the local complex viscoelasticity of the layer of ice and slush and is not to be confused with $\nu^*$. 

The quasistatic limit simplifies \eqref{momentum} through $\rho\ddot{u}=-\rho\omega^2u\approx 0$ and given the boundary conditions in \cite{Wang2010} it can be shown that $\nabla \cdot PI \approx \rho  g e_z\,$ \cite{Christian_thesis}. With these assumptions and relations in place, we arrive at the following equation and constitutive relation 
\begin{align} \label{system}
		\nabla \cdot \boldsymbol{\sigma}_s =0\,, 
		\quad
		\boldsymbol{\sigma}_s=2\nu\boldsymbol{\epsilon}_s \,.  
\end{align}
In a more general setting,  the stress and strain tensors are related through the fourth rank stiffness tensor $\boldsymbol{C}$ via $\boldsymbol{\sigma}=\boldsymbol{C} : \boldsymbol{\epsilon}$, where $(:)$ denotes contraction. Due to the incompressibility condition, $\boldsymbol{\epsilon}_v = 0$ and consequently $\boldsymbol{\epsilon}=\boldsymbol{\epsilon}_s$. However, this does not imply the same conditions hold for $\boldsymbol{\sigma}$, $\boldsymbol{\sigma}_v$, and $\boldsymbol{\sigma}_s$.


\subsection{Stieltjes integral representation}

We now depart from established continuum and rheological models and utilize the ACM to derive an alternative representation of $\nu^*$. 
The method explicitly incorporates the 
area fraction of the ice floe and slush/water mixture, and \emph{homogenizes} 
the equations in \eqref{system} to define a homogeneous
layer with scalar effective complex viscoelasticity 
$\nu^*$ that behaves energetically as the inhomogeneous layer on large scales. A key aspect of the method is providing a Stieltjes integral representation for $\nu^*$ involving the spectral measure of a self-adjoint operator which depends only on the geometry of the composite microstructure and the viscoelasticities $\nu_1$ and $\nu_2$ of the ice and water/slush constituents.
Examples of relevant geometries in this setting include the sea ice 
concentration, floe size distributions, and 
spacing between floes. 
We remark here that continuum models for the ice-ocean layer are inherently limited in terms of capturing the various wave-ice dynamics occuring in the MIZ; complex dynamical processes such as wave-induced drift, collision effects between moving ice floes, wave breaking, eddie currents and turbulence, etc. may not be amenable to such models and hence these shortcomings are necessarily present in the current model. As a result, the core parameters in continuum models are commonly tuned to help compensate for these more complicated processes. This will be discussed further in section \ref{Results}. 

Taking into account the heterogeneity of the layer, the stiffness/elasticity tensor $\textbf{C}(x)$ can be written in terms of the material phases as $\textbf{C}(x)=\textbf{C}_1 \chi_1(x) + \textbf{C}_2 \chi_2(x)$, where $\textbf{C}_1$ and $\textbf{C}_2$ are the stiffness tensors corresponding to the ice and water/slush, respectively. $\chi_1$ is an indicator function, taking the value 1 in the first phase of the material (ice) and 0 otherwise (water/slush), with $\chi_2=1-\chi_1$. The effective complex elasticity tensor $\textbf{C}^*$ of the ice-slush composite can be defined as \cite{milton2002,Cherkaev:JBiomech-345, Cherkaev:2019}
\begin{align}\label{eq:homogenized}
	\langle \boldsymbol{\sigma} \rangle=\textbf{C}^* : \langle \boldsymbol{\epsilon} \rangle .
\end{align}
Here, $\boldsymbol{\sigma}$ and $\boldsymbol{\epsilon}$ are 
stationary random fields and $\langle \cdot \rangle$ denotes ensemble averaging over all possible locally isotropic geometric realizations (or equivalently 
a spatial volume average via an ergodic theorem). We define the average deviatoric stress and strain as $\langle \boldsymbol{\sigma}_s \rangle=\boldsymbol{\sigma}_s^0$,  $\langle \boldsymbol{\epsilon}_s\rangle= \boldsymbol{\epsilon}_s^0$ and taking the deviatoric projection of equation (\ref{eq:homogenized}) defines the effective viscoelasticity of the homogenized layer as
\begin{align}
		\langle \boldsymbol{\sigma}_s\rangle=2\nu^* \boldsymbol{\epsilon}_s^0 ,
		\quad
		\langle \boldsymbol{\epsilon}_s \rangle = (1/2\nu^*) \boldsymbol{\sigma}_s^0\,.
\end{align}

The key step in the ACM is obtaining the following Stieltjes integral representation for $\nu^*$ \cite{Christian_thesis},
\begin{align}\label{intreps}
		\frac{\nu^*}{\nu_2} &=||\boldsymbol{\epsilon}_s^0||^2( 1-F(s))\,, 
		\quad  F(s)=\int_0^1 \frac{\text{d}\mu(\lambda)}{s-\lambda}\,,  
\end{align}
where $s=1/(1-\nu_1/\nu_2)$ represents the contrast between both material phases and the norm $||\cdot||$ is associated with the inner product $\langle f,g\rangle=\langle f : \bar{g} \rangle\
$ given by $||f||=\sqrt{\langle f,f\rangle}\,$, with $\bar{g}$ denoting complex conjugation. Equation \eqref{intreps} follows from the resolvent formula for the strain, derived from equation \eqref{system} \cite{Golden1983,Cherkaev:JBiomech-345,Murphy2015},  
\begin{align}\label{resolvents}
		\boldsymbol{\epsilon}_s=s(sI-\Gamma\chi_1)^{-1} \boldsymbol{\epsilon}_s^0\,,  
		\quad \Gamma =\nabla^S(\nabla \cdot \nabla ^S)^{-1} \nabla \cdot\, 
\end{align}
%
The random operator $\Gamma\chi_1$ is self-adjoint with respect to the above inner-product weighted by $\chi_1$ \cite{Cherkaev:JBiomech-345}, and $\Gamma$ is a projection operator onto the range of the symmetric gradient $\nabla^S$. Furthermore, the measure $\mu(\lambda)$ appearing in \eqref{intreps} is the \emph{spectral measure} associated with the random operator $\Gamma \chi_1$ \cite{Murphy2015}. The composite geometry is distilled into the measure through its moments
\cite{Bergman:PRL-1285,Milton:APL-300,Golden1983,milton2002,Cherkaev:2001,Cherkaev:IP-065008,Murphy2015}, 
e.g., the mass of the measure $\mu$ is given 
by $\mu^0=\langle\chi_1\rangle=\psi$, 
the sea ice concentration, and higher moments $\mu^n$ depend on the $(n+1)$-correlation functions of the random medium \cite{Golden1983,Bruno:JSP-365}. An important aspect of \eqref{intreps} is the separation of geometric effects in $\mu$ and constituent material parameters in $s$.

\subsection{Elementary Bounds} 
\label{elem_bounds}
%
Rigorous bounds for $\nu^*$ obtained from the integral in \eqref{intreps} provides quantitative information about $\nu^*$ from knowledge of the measure moments. As stated before, the measure mass is given by $\mu^0=\langle\chi_1\rangle=\psi$, and higher order moments are given by
\begin{align}
	\mu^n= \int_0^1 \lambda^n d\mu(\lambda)
	= \langle [\chi_1\Gamma\chi_1]^n 
	\chi_1 \hat{\boldsymbol{\epsilon}_s}^0: \hat{\boldsymbol{\epsilon}_s}^0 \rangle , \quad n=1,2,...
\end{align}
where $\hat{\boldsymbol{\epsilon}_s}^0=\boldsymbol{\epsilon}_s^0/||\boldsymbol{\epsilon}_s^0||$.

The elementary bounds are obtained from knowledge of $\mu^0$ and the constituent material properties $\nu_1, \ \nu_2$ alone. Following the procedure in 
\cite{Bergman:PRL-1285,Milton:APL-300,Golden1983,milton2002,Golden1986,Murphy2015}, we find $\nu^*$ must lie between the circular arcs in the complex $s$-plane,
\begin{align}\label{ebounds}
		Q= \nu_2 \left(1-\frac{\psi}{s-\lambda}\right),
		\quad
		\hat{Q}=\nu_1\bigg/ \left(1-\frac{1-\psi}{s-\hat{\lambda}}\right),
\end{align}
where $0\leq \lambda \leq 1-\psi$ and $0\leq \hat{\lambda} \leq \psi$ --- analogous to the elementary bounds for 
complex permittivity
\cite{Bergman:PRL-1285,Milton:APL-300,Golden1983,milton2002}.

\section{Data}
\label{data}

The dynamic region of the Arctic MIZ features multiple types of ice conditions, each exhibiting distinct wave-ice attenuation characteristics. In this section, we briefly detail a selection of published wave-ice laboratory and field datasets which are plotted in Section 4 along with the elementary bounds developed in Section \ref{elem_bounds}. Due to the multiple laboratory studies of wave dissipation in each type of ice condition, we group the laboratory datasets into three categories: (1) grease and grease-pancake ice, (2) broken floe field and pancake ice, and (3) continuous ice cover. The fourth category of data (4) is taken from field experiments which are a combination of multiple ice cover types.

\subsection{Grease and grease-pancake ice}
\label{grease-ice-section}
In \cite{newyear1997}, two sets of experiments measuring wave properties using five independent strain gauge probes were conducted on grease ice with thicknesses of $11.3$cm and $14.6$cm. We use data directly from Table 1 and Table 2 of \cite{newyear1997} and concatenate them into a single dataset. In \cite{Wang2010}, experiments were conducted in two parallel flumes (referred to as Tank 2 and Tank 3) during the REduced ice Cover in the ARctic Ocean (RECARO) project. Both flumes contained a mixture of grease and pancake ice with a varying thickness around $9$cm. The data from both Tank 2 and Tank 3 are given in Table 1 and Table 2 of \cite{Wang2010}, respectively, and are concatenated into a single dataset here. The wave experiments in \cite{Zhao2015} were conducted in Tank 3 (the same basin as \cite{Wang2010}) and we specifically utilize the data from Test 1 in \cite{Zhao2015} which was collected using a $2.5$cm thick frazil/pancake ice mixture. 

In \cite{YIEW2019256}, the attenuation and dispersion of waves were studied in a variety of ice covers in a wave flume using seven ultrasound sensors. We include data from two  grease ice experiments (with concentrations of $30$\% and $40$\%, concatenated into a single dataset) and a $12$cm thick grease-pancake ice mixture. Similarly, \cite{Parra2020} conducted wave studies on different ice covers at the Sea-Ice-Wind-Wave Interaction Facility at the University of Melbourne using seven ultrasound sensors in a wave flume. We use data from two grease ice experiments corresponding to $25$\% and $60$\% ice concentrations (not concatenated). 

Finally, we use the data produced by \cite{Rabault2019}, in which an array of six ultrasonic probes and particle image velocimetry sensors were combined in a small-scale wave tank experiment to study wave attenuation by grease ice with a thickness of $4$cm. The wavenumbers reported in \cite{Rabault2019} comes from a cross-correlation analysis of adjacent ultrasonic gauges. We use the wavenumbers obtained from the first two gauges (represented by $k_{12}$ in \cite{Rabault2019}).

\subsection{Broken floe field and pancake ice}
\label{broken-floe-section}
For this ice cover, we feature datasets collected in \cite{Zhao2015} from Test 2 and Test 3 which involves $4$cm thick pancake ice and a $7$cm thick broken floe field, respectively. From \cite{YIEW2019256}, we include datasets from a fragmented cover with a thickness of $4$cm and wide pancake ice with a thickness of $5$cm. Additionally, we include wave data from \cite{Parra2020} of ice floes that are $0.5$cm thick.

\subsection{Continuous ice cover}
\label{continuous-section}
Laboratory datasets for continuous ice cover primarily come from \cite{YIEW2019256}. Four small datasets are provided here, featuring continuous ice cover with thicknesses of $0.5$cm, $1$cm, and $1.5$cm as well as cemented pancake ice. One more dataset from \cite{Parra2020} of $0.5$cm thick long sheet ice is also included.

\subsection{Field observations}
\label{field-section}
In \cite{Kohout2014}, strong correlation between wave activity and sea ice extent was found using in-situ measurements of wave attenuation on a track inward along the MIZ during the Sea Ice Physics and Ecosystem Experiment II (SIPEXII) Antarctic cruise in 2012. In \cite{Meylan2014,Meylan2018}, an empirical relationship between wave attenuation and period was found which displays power law attenuation behavior. The energy attenuation rate of waves in the MIZ with ice fractions between 20\%-60\% follow the relationship
\begin{align}\label{mikes}
	k_i=\frac{a}{T^2}+\frac{b}{T^4}, \quad a=2.12\times10^{-3}, \ b=4.59\times10^{-2} .
\end{align}

In \cite{Wadhams1988} field observations by the Scott Polar Research Institute (SPRI) between 1978 and 1983 in the Greenland and Bering Seas are analyzed. Several experiments were conducted using wave buoys and accelerometer packages installed on floes to measure wave energy along a line of stations extending from open sea to the interior dense ice pack. While \cite{Wadhams1988} features multiple experiments, we only use the datasets from the 1979 Greenland Sea: September 4 and the 1983 Bering Sea: February 7 experiments as  recommended in \cite{KohoutMeylan2008, XuGuyenne}. In both the data from \cite{Wadhams1988} and the attenuation rule \eqref{mikes}, the ice conditions are characterized by large broken floes amongst an oceanic host. The attenuation in both these data sets are given in terms of wave \emph{energy}. Since the current model involves the wave amplitude attenuation, we account for this disparity by multiplying the attenuation data by a factor of $1/2$.

\subsection{Plotting $\nu^*$ from data}
In \cite{Zhao2015, Cheng2017, XuGuyenne}, laboratory measurements of the frequency and complex wavenumber $(\omega, k)$ are plugged into the dispersion relation of \cite{Wang2010} which enables the effective roots $(G, \upsilon)$ to be obtained via an optimization procedure. A similar but simpler approach is to choose values $(G, \upsilon)$ such that the resulting complex wavenumber exhibits an attenuation rate that most agrees with observations. However, these approaches lead to complications, namely, a large number of pairs $(G,\upsilon)$ ultimately yield attenuation rates that match experiments \cite{Mosig2015} and specific assumptions must be taken into account in order to choose the most physically appropriate $(G, \upsilon)$ pair. Due to the quasistatic assumption of our model, the dispersion relation \eqref{SDP} is substantially simpler than other continuum models and avoids these difficulties. In the quasistatic setting, frequency and complex wavenumber data $(\omega, k)$ can be plugged into equation \eqref{SDP} to determine \emph{unique} values of $\nu^*$ which itself is a representation of the effective parameters $(G, \upsilon)$. This is the approach taken in the following numerical results. For both field datasets, wavenumbers $k_r$ are not provided and so we assume an open water wavenumber satisfying the dispersion relation $\omega^2 = gk_r\tanh{Hk_r}$ where the depth of water $H$ is taken to be $1500$ m. The wavenumber in ice-covered seas is observed to be approximately the same as the open water wavenumber, and so this choice is suitable for the current setting. For all datasets, we only use the average values of $(\omega, k)$ provided, ignoring the reported standard deviation errors.

\subsection{Constituent parameters for elementary bounds}

To numerically compute the bounds for $\nu^*$ in \eqref{ebounds}, values of the constituent visoelasticities for ice $\nu_1=G_1-i\rho\omega\upsilon_1$ and water/slush $\nu_2=G_2-i\rho\omega \upsilon_2$ are needed. In particular, we are faced with the nontrivial task of choosing a suitable range of established values for $(G_1, v_1)$ and $(G_2, v_2)$ to achieve physically accurate bounds for multiple types of ice covers. The elastic modulus of sea ice has been estimated to be around $1$ GPa \cite{Timco2010} ,  $0.4-0.7$ GPa \cite{voermans_2023} , and $0.77$ GPa \cite{Marchenko2013}, and the dynamic viscosity of sea ice can vary between $10^8-10^{13}$ Pa$\cdot$s \cite{Voermans2021}. In \cite{YIEW2019256}, the dynamic viscosity of a $40\%$ concentration of grease ice was directly measured to be $0.12 \pm 0.05$ Pa$\cdot$s, however, the measured wave attenuation data fit to the model of Wang and Shen instead yielded  values of  $10-50$ Pa$\cdot$s. The elastic modulus and kinematic viscosity of water can be taken to be $0$ Pa$\cdot$s and $10^{-6} \ \text{m}^2 \text{s}^{-1}$, respectively. For simplicity, we consider only a homogenized density $\rho=974 \text{ kg{/}m}^3$.

Choosing $(G_1, v_1)$ and $(G_2, v_2)$ leads to other physical complications, for example, water is not typically characterized by an elastic modulus $G_2$, which is a measure of the stiffness of a solid material. Grease ice, a homogeneous mixture of frazil ice crystals and supercooled water, cannot be described as a geometrically separated two-phase material in the same way that a broken floe field can, leading to difficulties in interpreting the roles of $(G_1, v_1)$, $(G_2, v_2)$, and $\psi$. 
For sea ice and water, the differences between the parameters  $(G_1, v_1)$ and $(G_2, v_2)$ can exceed $10$ orders of magnitude, leading to extremely large bounds and rendering them unfit for physical interpretations. The estimated effective values of the shear modulus $G$ and kinematic viscosity $\upsilon$ for the homogenized ice layer must account for various processes, including dissipation effects due to scattering, eddy viscosity, collision between ice floes, etc., and thus can vary by several orders of magnitude depending on both ice conditions and the specific ice-ocean model being used \cite{XuGuyenne}. Finally, we acknowledge that the datasets showcased here commonly feature wave frequencies higher than $1$ Hz which is incongruent with the quasistatic assumption of our model, e.g., the scattering of higher frequency waves which affect the attenuation in each dataset cannot be accounted for with the current constituent equations and dispersion relation. These factors together lead to a practical compromise of using a simple fitting process to reach final values of  $(G_1, v_1)$ and $(G_2, v_2)$ for each ice cover type -- a common tactic for enhancing the predictive abilities of continuum models.

In addition to the constituent parameters given above, \textit{effective} parameters of floating ice covers have been measured and estimated in both laboratory and field settings. In \cite{Marchenko2021}, the effective elastic modulus was measured to be $46-378$ MPa. Table $1$ in \cite{XuGuyenne} offers a comprehensive list of effective parameters inversely computed using three different wave-ice models and multiple datasets, finding the effective elastic modulus $G$ to take values in the range  $6.40 \times 10^2 - 6.50 \times 10^{11}$ Pa and the effective kinematic viscosity $v$ to take values in the range $1.22 \times 10^{-4} - 4.62 \times 10^{7}$ $\text{m}^2\text{s}^{-1}$. Recall that the kinematic viscosity $v$ and dynamic viscosity $\mu$ are related via $\rho v = \mu$.

\section{Results and discussion}\label{Results}



To choose the constituent parameters $(G_1, v_1)$ and $(G_2, v_2)$, we perform a search over the values $10^2 \leq G_1, G_2 \leq 10^9$ Pa$\cdot$s and $10^{-4} \leq v_1, v_2 \leq 10^8$ $\text{m}^2\text{s}^{-1}$ which reasonably encompass the ranges reported for both constituent and effective parameters given above. Each combination of constituent parameters corresponds to unique elementary bounds. The final values for $(G_1, v_1)$ and $(G_2, v_2)$ are chosen such that they produce the smallest bounds which still contain all of the data of a given ice cover dataset. Since ice area fractions are seldom recorded along with wave attenuation data, we choose values for $\psi$ that are physically reasonable for each type of ice cover. Finally, the wave period $T$ is taken to be the average of periods of all the data points collected for each ice cover dataset. 

\begin{figure}[]              
	\centering
	\includegraphics[width=\textwidth]{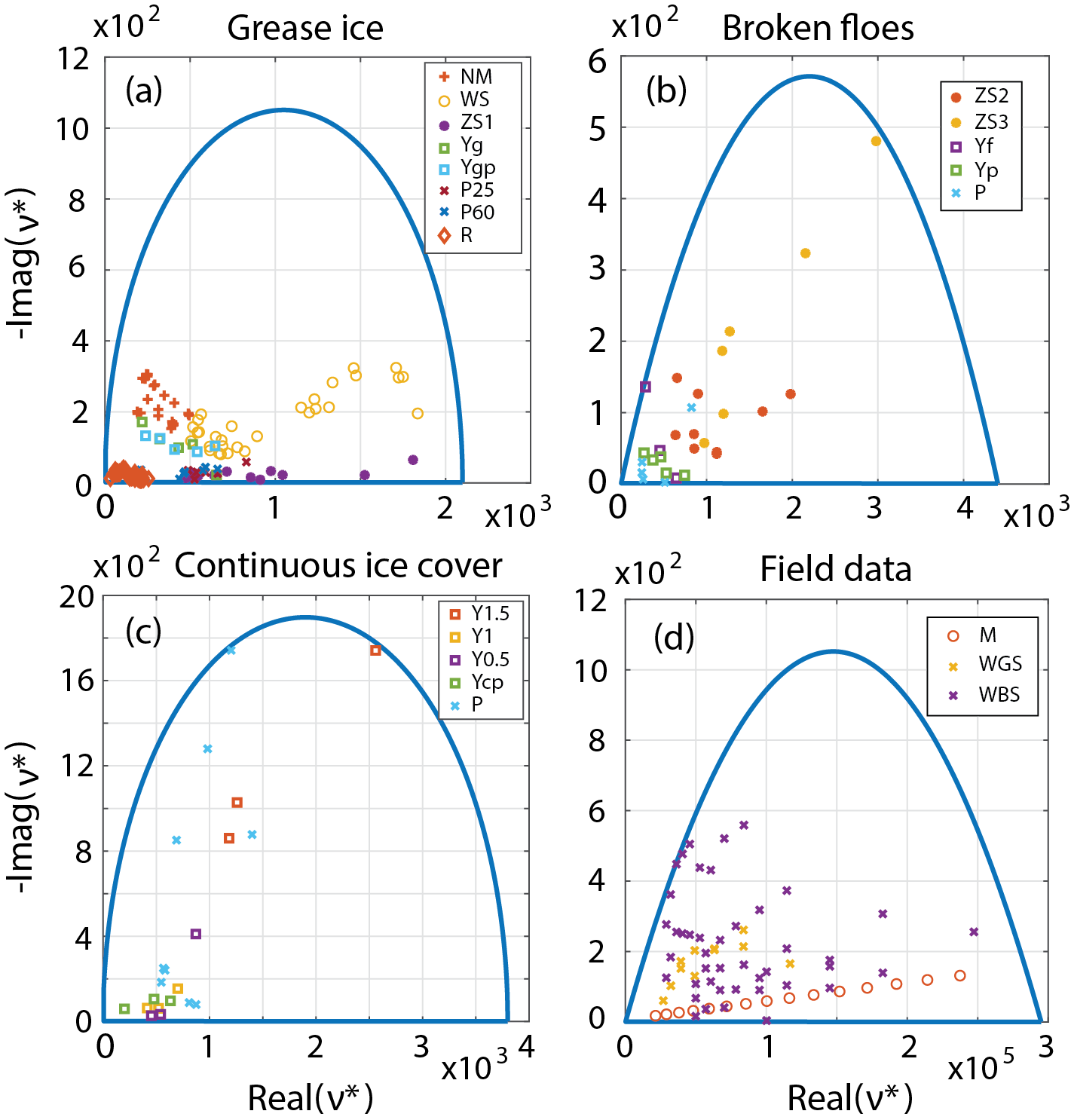}
	\caption{Frequency, wavenumber, and attenuation data $(\omega, k)$ interpreted through equation (\ref{SDP}) with fitted elementary bounds for different ice covers. (a) Grease ice and grease-pancake ice data taken from \protect\cite{newyear1997} (NM),  \protect\cite{Wang2010} (WS), \protect\cite{Zhao2015} Test 1 (ZS1), \protect\cite{YIEW2019256} (Yg (grease) and Ygp (grease-pancake), \protect\cite{Parra2020} (P25 and P60 corresponding to $25\%$ and $60\%$ concentrations of grease ice), and \protect\cite{Rabault2019} (R). (b) Broken floe and pancake ice data taken from \protect\cite{Zhao2015} Test 2 and Test 3 (ZS2 and ZS3), \protect\cite{YIEW2019256} (Yf (fragmented floes) and Yp (wide pancakes)), and \protect\cite{Parra2020} (P). (c) Continuous ice cover data taken from \protect\cite{YIEW2019256} with thicknesses 1.5cm, 1cm, and 0.5cm (Y1.5, Y1, and Y0.5), cemented pancake ice (Ycp), and \protect\cite{Parra2020} (P). (d) Field data taken from \protect\cite{Wadhams1988} Greenland Sea and Bering Sea (WGS and WBS), and the attenuation rule (\ref{mikes}) from \protect\cite{Meylan2014} (M) corresponding to $T=6\text{s}$ (leftmost point) and extending to $T=20\text{s}$ (rightmost point) in increments of $1\text{s}$. }
	\label{fig:grease_ice}
\end{figure}

Figure \ref{fig:grease_ice}a displays the data points collected for grease ice and grease-pancake ice described in Section \ref{grease-ice-section}.  We take $T=0.66$s and $\psi=0.42$ which are averages of the reported wave periods and ice concentrations. The tuned parameters are found to be $G_1=5000$ Pa$\cdot$s, $G_2=10^{-6}$ Pa$\cdot$s, and $v_1=v_2=10^{-4} \ \text{m}^2 {/} \text{s}$. The broken floe field and pancake ice dataset of Section \ref{broken-floe-section} is shown in Figure \ref{fig:grease_ice}b. For the bounds we let $T=1.1$s and $\psi=0.55$. The tuned parameters for this ice cover are found to be $G_1=8000$ Pa$\cdot$s, $G_2=1$ Pa$\cdot$s, and $v_1=v_2=10^{-4} \ \text{m}^2 {/} \text{s}$. Figure \ref{fig:grease_ice}c displays the data and bounds for continuous ice covers described in Section \ref{continuous-section} with $T=0.926$ seconds, $\psi=0.95$, and tuned parameters $G_1=4000$ Pa$\cdot$s, $G_2=10^{-4}$ Pa$\cdot$s, $v_1=10^{-4} \ \text{m}^2 {/} \text{s}$, and $v_2=10^{-5} \ \text{m}^2 {/} \text{s}$. Finally, Figure \ref{fig:grease_ice}d displays the data and bounds for the field data described in Section \ref{field-section}. For the bounds, we let $T=10.47\text{s}$, $\phi=0.5$, and tuned parameters $G_1 = 5.9\times 10^5$ Pa, $v_1=10^{-4} \ \text{m}^2 {/} \text{s}$, $G_2=41$ Pa, and $v_2=10^{-3} \ \text{m}^2 {/} \text{s}$.

The results in Section 4 reveal unique effective parameters when tuning the model to fit sea ice laboratory and field wave data. The fitted elastic modulus of sea ice $G_1 \sim 10^3 - 10^5$ Pa is several orders of magnitude lower than the typical range reported in the literature. Meanwhile, the elastic modulus of water/slush $G_2 \sim 10^{-6}-10^1$ can be considered to be more realistic. The kinematic viscocity $v$ for ice and ocean phases are both in the range $10^{-5} - 10^{-3} \ \text{m}^2 {/} \text{s}$  which is physically reasonable for water but not for ice. As discussed before, it is expected to recover non-physical material properties when fitting models to a comparatively complex system such as waves propagating through sea ice. Such characteristics are already present when fitting parameters to the full dispersion relation of \cite{Wang2010} and are certainly exaggerated by the quasistatic assumption of the current model. The current model explicitly accounts for wave attenuation due to the energy storage of each phase (e.g. bending of ice floes which is controlled by elasticity) and dissipation due to friction (controlled by viscosity). Other dynamic processes such as wave scattering, inelastic bumping and deformation of floes, drag of floes through viscous slush, waves washing over floes, etc. are invisible to the current model. However, the characteristics and attenuation due to these processes are still present in the data, leading to necessary compensation through the material parameters.

We hypothesize that the model compensates by requiring the elastic modulus of ice to be quite small, allowing for more wave attenuation to result from elastic deformations of the ice phase. At first glance, the small values for the viscosity of ice suggest that elastic deformations dominate the wave attenuation, however, the axes in Figure \ref{fig:grease_ice}(a), (b), and (c) illustrate that the imaginary part of $\nu^*$ lags behind the real part by only one order of magnitude (recall that $\text{Re}(\nu^*)$ and $-\text{Im}(\nu^*)$ correspond to effective elasticity and viscosity, respectively). This trend is broken by the field data in Figure \ref{fig:grease_ice}(d), where $\text{Re}(\nu^*)$ exceeds $-\text{Im}(\nu^*)$ by up to three orders of magnitude. There is a positive correlation between $\text{Re}(\nu^*)$ and $-\text{Im}(\nu^*)$ for broken floes, continuous cover, and field data which suggests that as the layer becomes stiffer it also tends to exhibit more viscous behavior. This trend is absent in grease ice which maintains a comparably more steady viscosity across a range of elasticities.  The data derived from the attenuation rule (\ref{mikes}) illustrates that, as wave periods grow larger, elasticity plays an increasingly larger role than viscosity for describing the homogenized, macroscopic behavior. In general, the field data exhibits larger values of elasticity over the other laboratory ice types (and hence a larger fitted $G_1$ value), suggesting that the complex mixture of ice types and influences from field experiments are more adequately accounted for by increased elasticity of the ice phase.

\section{Conclusion}
We have developed a novel approach for analyzing wave attenuation in ice-covered seas in a variety of different rheological settings. This approach provides rigorous bounds on the effective complex viscoelasticity of the ice-ocean composite based purely on sea ice concentration and constituent material properties. Comparison with experimental data is achieved through a quasistatic dispersion relation. The bounds successfully capture a diverse range of laboratory and field wave attenuation measurements across different ice morphologies and spatial scales. While the fitted material parameters deviate from typical values for solid sea ice - likely due to the model's quasistatic assumptions and simplified treatment of complex wave-ice interactions - the bounds provide valuable constraints on effective parameters used in operational wave-ice models. 
Proper, non-arbitrary constraints on the effective viscoelasticity can help to validate the tuned values of more computationally robust models which can produce values of wildly varying magnitudes. 
Physically motivated bounds are a crucial element for efficient sampling strategies in measuring model behavior, e.g. uncertainty quantification. The ability of the bounds to encompass datasets from diverse ice morphologies and spatial scales hint at the universality of fundamental physical principles which may lie at the heart of future wave-ice models. This broad applicability is particularly noteworthy given the quasistatic assumption of the model, a simplification that offers the benefit of vastly streamlined parameterizations and computational efficiency. Furthermore, the bounds capture the essential physics and behavior of observed wave attenuation without any mathematical machinery which explicitly quantifies such processes. 
This framework offers a promising foundation for improving parameterizations of wave attenuation in sea ice models and advancing our understanding of the critical role waves play in the evolution of Earth's ice covers.

\section{Acknowledgments}
We gratefully acknowledge support from the 
Division of Mathematical Sciences at the US National Science 
Foundation (NSF) through Grants DMS-0940249, DMS-1413454,
DMS-1715680, DMS-2136198, and DMS-2206171.
We are also grateful for support from the Applied and 
Computational Analysis Program 
and the Arctic and Global Prediction Program 
at the US Office of Naval 
Research through grants 
N00014-13-1-0291,
N00014-18-1-2552,
N00014-18-1-2041
and
N00014-21-1-2909.
Finally, we would like to thank the NSF Math Climate Research Network (MCRN), and especially Chris Jones, for supporting this work, and providing funds for collaborative efforts abroad. 


\bibliographystyle{abbrv}  
\bibliography{WaveChapter,golden}  %

\end{document}